# Bayesian Optimization of Multi-Bit Pulse Encoding in $In_2O_3/Al_2O_3$ Thin-film Transistors for Temporal Data Processing

Javier Meza-Arroyo*[a], Benius Dunn[a], Weijie Xu[a], Yu-Chieh Chen[b], Jen-Sue Chen[b], Julia W.P. Hsu*[a]

[a] Department of Materials Science and Engineering, The University of Texas at Dallas, Richardson, Texas 75080, USA

[b] Department of Materials Science and Engineering, National Cheng Kung University, Tainan 70101, Taiwan.

*Correpondence to: J. Hsu (jwhsu@utdallas.edu); J. Meza (javiermezaa7@gmail.com)

## Abstract

Utilizing the intrinsic history-dependence and nonlinearity of hardware, physical reservoir computing is a promising neuromorphic approach to encode time-series data for in-sensor computing. The accuracy of this encoding critically depends on the distinguishability of multi-state outputs, which is often limited by suboptimal and empirically chosen reservoir operation conditions. In this work, we demonstrate a machine learning approach, Bayesian optimization, to improve the encoding fidelity of solution-processed $Al_2O_3/In_2O_3$ thin-film transistors (TFTs). We show high-fidelity 6-bit temporal encoding by exploring five key pulse parameters and using the normalized degree of separation (nDoS) as the metric of output state separability. Additionally, we show that a model trained on simpler 4-bit data can effectively guide optimization of more complex 6-bit encoding tasks, reducing experimental cost. Specifically, for the encoding and reconstruction of binary-patterned images of a moving car across 6 sequential frames, we demonstrate that the encoding is more accurate when operating the TFT using optimized pulse parameters and the 4-bit optimized operating condition performs almost as well as the 6-bit optimized condition. Finally, interpretability analysis via Shapley Additive Explanations (SHAP) reveals that gate pulse amplitude and drain voltage are the most influential parameters in achieving higher state separation. This work presents the first systematic method to identify optimal operating conditions for reservoir devices, and the approach can be extended to other physical reservoir implementations across different material platforms.

## Introduction

The rapid development of artificial intelligence and the Internet of Things is driving a pressing demand to overcome the latency and energy overheads associated with constantly shuttling data between memory and logic circuits[1]. Edge computing strategies that integrate memory and processing in one platform based on brain-inspired computing architectures, particularly neuromorphic systems, have been growingly explored to achieve real-time, energy-efficient computation to overcome the limitations of traditional von Neumann computing architectures[2,3].

Processing temporal data is essential for applications such as time-series classification, speech recognition, and sensory preprocessing[4–6]. Among various neuromorphic approaches, physical reservoir computing (PRC) uses intrinsic history-dependence and nonlinearity in hardware to encode temporal data. Recent advances in hardware implementations of PRC have explored a variety of device platforms, such as memristors[7,8], thin film transistors (TFTs)[9–11], and electrochemical transistors[12,13]. Unlike two-terminal memristors, which often require an additional access transistor to mitigate sneak-path currents in crossbar arrays[14], three-terminal oxide TFTs enable independent control of internal state modulation and signal readout through separate terminals, thereby suppressing electrical crosstalk. Moreover, solution-processed oxide TFTs offer additional advantages, including compatibility with large-area fabrication, flexible substrates, and low-temperature processing. Notably, certain oxide TFTs exhibit intrinsic hysteresis and time-dependent charge dynamics, allowing them to emulate short-term memory effects similar to those observed in biological synapses [15]. These characteristics can be exploited to construct reservoirs in which transient electrical states represent past inputs, effectively enabling in-memory computation.

Upon being excited by input stimuli, a reservoir dynamically generates a high-dimensional state space, which is subsequently processed by a simple readout layer. Because the internal parameters of the reservoir remain fixed, the training complexity and cost of PRC is much lower[16,17]. The richness associated with its nonlinearity and dynamics enables the reservoir to generate distinct responses for similar time-series signals with different histories, ideal for the encoding of temporal data[18–20]. The effectiveness of PRC is thus critically dependent on operating the reservoir at a condition that produces well distinguishable outputs for different time-sequence data, especially when aiming for high-bit operation with many output states.

Thus, a key challenge in using TFTs for temporal data processing is to identify the optimal operating condition. Pulse-based input encoding is a common technique where sequences of voltage pulses modulate the internal state of the transistor. For 4-bit encoding, there are only 16 distinct pulse sequences, so conventional methods by trial-and-error, varying one variable at a time, or based on the researcher's experience are able to obtain a good operation condition, as demonstrated in several oxide-based TFT systems[13,15,21–23]. Yet, scaling this approach to 5, 6, or even more bits of encoding, requiring more than 32, 64, and more well-separated output states, is difficult to manage based on the tour de force approach. Additionally, very few reports—only two to our knowledge[4,25]—explicitly quantify the distinguishability of output states using a defined metric. Previous demonstration of 5-bit operation reported promising classification accuracies, but the output state distributions exhibited significant clustering and poor uniformity even at 4 bits[24]. Similarly, Park et al.[9] reported a chaotic and poorly distributed current response in 5-bit encoding experiments. Even more advanced demonstrations, such as the 8-bit encoding by Dongyeol et al.[7], showed noticeable aggregation of states, particularly for sequences ending in '0', which undermines state distinguishability. These observations highlight how the inherent nonlinearity and hysteretic, history-dependent nature of TFT behavior pose a high-dimensional, complex optimization problem. Small variations in pulse amplitude, duration, or spacing can lead to substantial changes in the output signals, and identifying the optimal combination of these parameters through direct experimentation becomes prohibitively time- and resource-intensive[17]. Hence, to advance the adoption of PRC, there is a need for a systematic approach that is based on quantitative evaluation of output distinguishability and does not rely on the operator's intuition or experience.

In this work, we demonstrate the use of Bayesian optimization (BO), a machine learning framework designed to efficiently explore high-dimensional input (feature) spaces under limited experimental budgets[26,27]. BO constructs a probabilistic model of the objective function with the known experimental data using Gaussian process regression (GPR) and suggests new experiments based on an acquisition policy that balances exploration and exploitation[28]. The active learning process iterates until a convergence is achieved. In this work, we apply BO to systematically optimize five pulse parameters -- pulse period, base gate voltage, gate pulse amplitude, drain voltage, and duty cycle -- in a sol–gel processed $In_2O_3/Al_2O_3$ TFT, to maximize the separation among the 64 output states generated from 6-bit binary sequences. For the performance metric to guide the optimization, we use the normalized degree of separation (nDoS), which is similar to degree of separability[25] or DS ratio[4] proposed previously. We validate our optimized 6-bit operating condition by applying it to a motion image processing task, demonstrating the practical utility of our approach for real-world spatiotemporal signal analysis. This ability to produce well-separated output states serves as a proxy for the reservoir's capacity to distinguish between different temporal patterns, a critical requirement for reliable neuromorphic computation.

Since optimizing 6-bit operation condition is time-consuming and complex, we investigate using a 4-bit encoding model as a surrogate to guide the optimization of higher-bit data processing. By training GPR models in parallel on both 4-bit and 6-bit pulse sequences, we assess the degree of correlation between their optimized parameter spaces and evaluate whether the simpler 4-bit model can effectively predict high-performing conditions for 6-bit encoding tasks. This strategy has the potential to significantly reduce experimental overhead while providing a scalable pathway for implementing high-bit neuromorphic encoding. We show that, for a 6-bit motion image encoding task, using the optimized 4-bit operating condition performs similarly to using the directly optimized 6-bit operating condition. Finally, we perform a Shapley Additive Explanations (SHAP)-based interpretability analysis to quantify the contribution of each pulse parameter to the predicted state separation, providing insight into the relative importance and interactions of the input variables[29,30]. This analysis identifies the variables that most strongly influence the encoding performance, further guiding future optimization and device design.

Taken together, this work offers a novel machine learning-assisted design methodology for optimizing high-bit encoding for temporal data processing. By leveraging hysteresis and nonlinear dynamics in oxide semiconductors and coupling them with data-driven optimization, we provide a compelling pathway toward scalable, energy-efficient, and intelligent hardware systems for next-generation neuromorphic computing.

**Bayesian Optimization of Pulsed Voltage Conditions**

Achieving high-resolution multi-bit encoding in physical systems requires precise control over pulse parameters that govern the transient dynamics of the TFT. However, the high dimensionality and nonlinear, history-dependent behavior of these devices make manual optimization inefficient and often suboptimal. To address this, we implemented a BO framework that enables data-efficient exploration of the parameter space and rapid convergence toward optimal operating conditions.

Fig. 1 illustrates the overall optimization workflow. The initial exploration of the five-dimensional inputs -- pulse period, base gate voltage, gate pulse amplitude, drain voltage, and duty cycle -- uses Latin hypercube sampling (LHS) to generate a set of 20 conditions that ensures a well-dispersed sampling across all input dimensions (Fig. 1(a)). These five parameters collectively define a vast search space with over 12 million possible combinations (operating conditions), making exhaustive or grid-based optimization approaches infeasible. The distribution of LHS-sampled values for each parameter and the parallel plot are shown in Fig. S1. Each LHS condition is applied to the neuromorphic TFT using a 6-bit pulse train (e.g., 111111), and the resulting time-dependent current is recorded (Fig. 1(b)). The current values following each pulse (highlighted with red markers) are extracted to construct normalized current vs. number of pulses plots for all 64 sequences. The final current from each sequence is used to calculate the nDoS (Methods), which quantifies how well-separated the output states are. Since nDoS values vary by many orders of magnitude, we use the logarithm of the value, log(nDoS), as the objective function to be maximized during the optimization process (Fig. 1(c)).

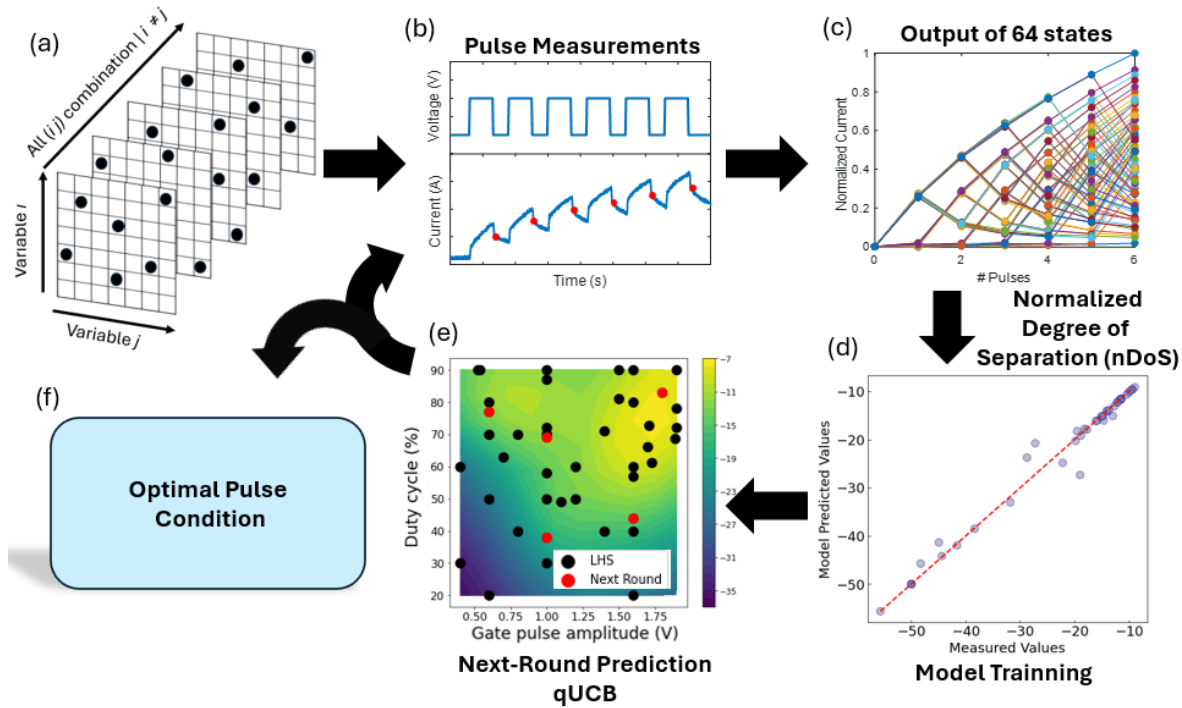

*Fig. 1 BO workflow for selecting pulse condition that maximizes the distinguishability among the outputs for different pulse sequences. (a) Initial experimental conditions generated by LHS. (b) Application of the 111111 pulse voltage waveform to the TFT (top) and the corresponding current transient (bottom). (c) Normalized current values for 64 pulse sequences for the determination of log(nDoS) value. (d) Training of the GPR model with measured nDoS values. (e) Prediction of the next-batch input conditions using qUCB acquisition function. (f) Identification of the pulse voltage condition that produces maximum nDoS value.*

A GPR model is trained on the initial data (Fig. 1(d)) and used to predict high-potential input conditions through a batch Upper Confidence Bound (qUCB) acquisition function (q = 5). See Methods for details. These candidates are experimentally evaluated (Fig. 1(e)), and the process is iteratively repeated, updating the model until convergence is reached. Convergence is defined as the absence of further meaningful improvement in

log(nDoS), at which point the pulse condition yielding the highest value is selected as the optimal configuration (Fig. 1(f)).

**Electric Characterization of TFT**

Fig. 2(a) illustrates the structure of the TFT, employing a bottom-gate/top-contact architecture. The device consists of a sol–gel derived $Al_2O_3$ dielectric layer and an $In_2O_3$ semiconductor channel. The transfer characteristics (drain current – gate voltage) measured under a gate voltage sweep from –1 V to +3 V at a fixed drain voltage ($V_{DS}$) of 3 V are shown in Figure 2(b), displaying five sweeps out of a total of eight performed. The device exhibits a highly repeatable counterclockwise hysteresis loop with a threshold voltage shift of approximately -0.5 V. This hysteresis is primarily attributed to gate-induced motions in oxygen vacancies ($V_O^+$) in the $Al_2O_3$ dielectric, as reported in previous works[31–33]. During the forward sweep toward larger positive gate voltage, $V_O^+$ are driven toward the dielectric/semiconductor interface. Their accumulation near the interface induces a localized electric field that increases electron density in the $In_2O_3$ channel, enhancing the drain current. In the backward sweep, due to the relatively slow relaxation kinetics of the $V_O^+$, a remnant field persists, resulting in a continued high electron density and a higher current compared to the initial value. This mechanism supports observed hysteresis and contributes directly to the device's short-term memory behavior. A closer inspection of the low-current regime, magnified in Figure 2(c), reveals a progressive negative shift in the threshold voltage with each sweep. This drift is attributed to cumulative charges, likely caused by incomplete relaxation of $V_O^+$ between measurements. The resulting electron accumulation at the channel interface slightly increases the drain current over successive sweeps, consistent with the dynamic hysteresis mechanism described previously.

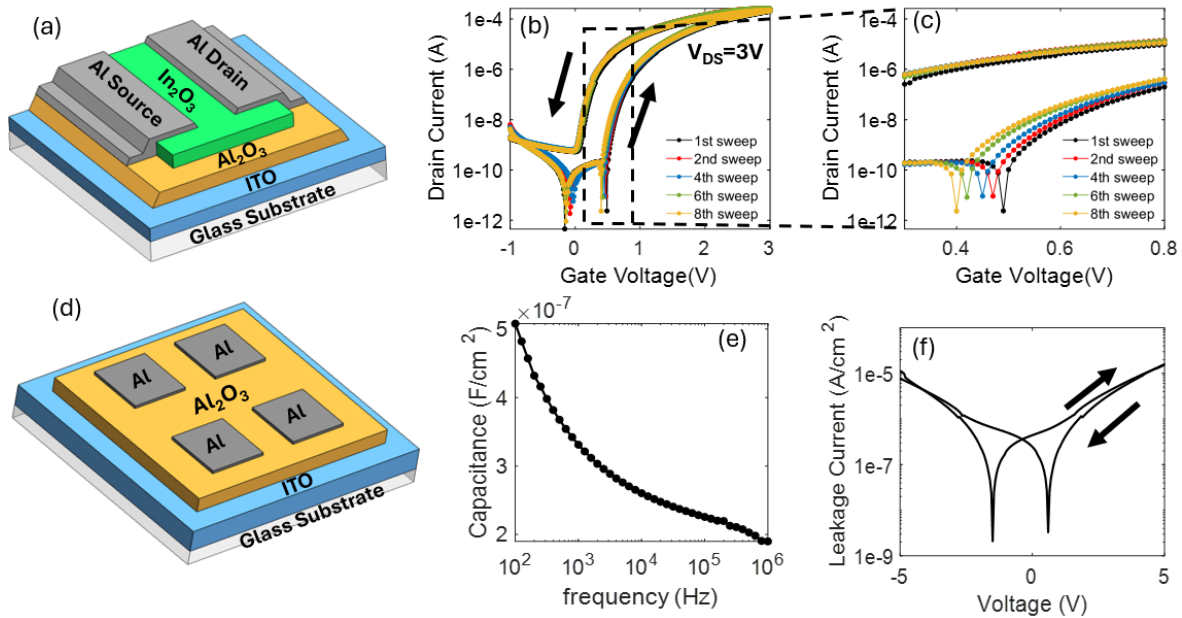

Fig. 2 (a) Schematic of the $In_2O_3/Al_2O_3$ TFT. (b) Transfer characteristics showing counterclockwise hysteresis with threshold voltage shift of -0.5 V. (c) Zoom In of transfer curves. (d) Schematic of MIM structure with $Al_2O_3$ as dielectric. (e) C-f and (f) I-V characteristics of MIM device.

However, this cumulative shift can be mitigated and the device restored to its initial state by applying a reset protocol consisting of a brief (3 s) negative bias stress ($V_G$ = -2 V, $V_{DS}$= -3 V). Figure S2 shows that the drain current increases with consecutive biases, but after the reset, all first bias measurements return to the same initial current level, with no memory of previous measurements. This reset protocol ensures a reproducible initial condition for subsequent measurements, which is essential for multi-bit encoding tasks.

To further validate the role of these defects in the dielectric behavior, capacitance vs frequency (*C–f*) measurements were performed on Metal-Insulator-Metal (MIM) capacitors (Fig. 2(d)) using the same $Al_2O_3$ dielectric. As shown in Figure 2(e), the capacitance is ~ 200 nF/cm² at 1 MHz but increases significantly to ~ 500 nF/cm² at 100 Hz. This pronounced frequency dependence indicates a slow polarization process, consistent with the response of polarizable species such as $Vo^+$, OH groups, or absorbed water in the $Al_2O_3$ dielectric that only contribute to the dielectric constant at low frequencies[34–36]. In addition, Figure 2(f) shows the leakage current–voltage (*I–V*) characteristics of the MIM capacitor, exhibiting a leakage current of $10^{-5}$ A/cm² at 5 V. Notably, the *I–V* curve also exhibits a hysteresis loop, further supporting the presence of mobile defects within the dielectric layer[32].

These results confirm that the $In_2O_3/Al_2O_3$ TFT not only exhibits robust hysteresis and memory behavior required for neuromorphic tasks, but also offers excellent repeatability and reconfigurability, making it a suitable candidate for multi-pulse measurements and memory-dependent computing operations. The ability to deliberately control the internal state of the device via electrical biasing is a powerful asset for physical reservoir computing systems that rely on consistent yet reconfigurable nonlinear dynamics.

**Six-bit Pulse Encoding and Output State Behavior**

To demonstrate the current response dynamics of the sol-gel $In_2O_3/Al_2O_3$ TFT under multi-bit pulses, we analyze the output behavior for a few examples of 6-bit binary sequences. Figure 3(a)-(c) summarizes this process using four representative pulse sequences: 111111, 001101, 011000, and 010000. In Figure 3(a), the applied gate voltage waveforms are shown for each of the four sequences. Each sequence consists of six binary pulses, where "1" corresponds to applying a high gate voltage pulse (base gate voltage + gate pulse amplitude) and "0" to applying the base voltage to the gate, as given by the selected input condition. The temporal structure of the pulses reflects the binary encoding used to represent temporal signals. The corresponding current responses are presented in Figure 3(b), where the current is plotted against time for each sequence. The dynamic evolution of the current under each pulse is visible, and the dot markers indicate the specific time points at which the current values are extracted to visualize the signal progression during the full pulse trains.

Figure 3(c) shows these extracted current values as a function of the pulse number for the four sequences. Each curve demonstrates a unique evolution of current with each pulse, reflecting the nonlinearity of the device and history-dependent behavior. The final current values in Figure 3(b) are used to represent the output state of the transistor after receiving a specific pulse sequence. As shown in Figure 3(c), the final current levels after six pulses are clearly distinguishable between the four sequences, illustrating the underlying principle behind multi-bit encoding.

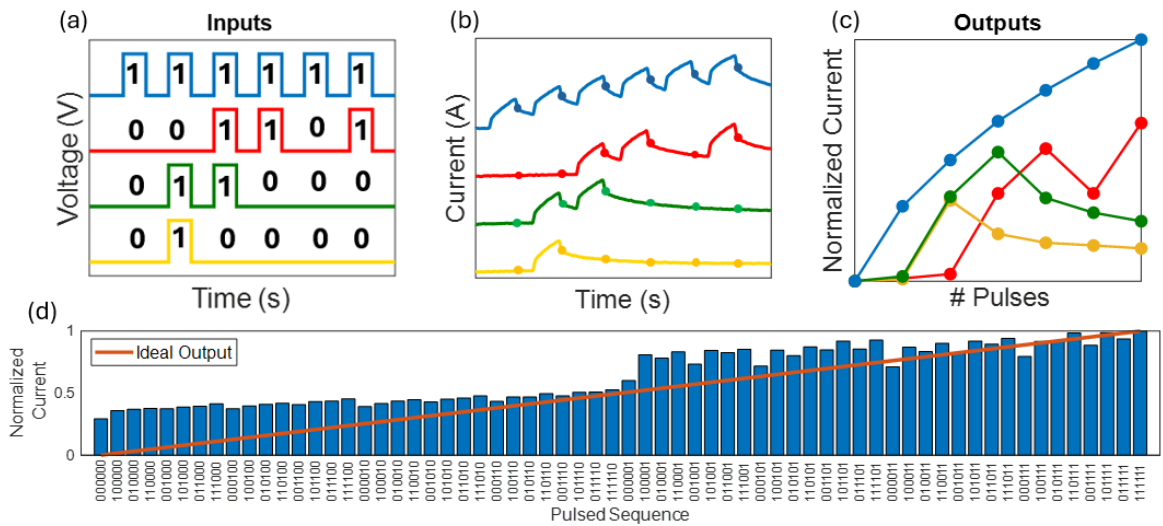

*Fig. 3 (a) Gate voltage pulse trains for four representative 6-bit binary sequences. (b) Source current response as a function of time during the corresponding pulse application. (c) Normalized output current vs # of pulses for the four*

*sequences, showing four distinguished outputs. (d) Comparing the ideal evenly spaced final result (red line) and experimental results obtained from an LHS condition (blue bars) for all 64 pulse sequences in 6-bit encoding.*

Finally, Figure 3(d) compares the output currents of all 64 possible 6-bit sequences using a bar plot for two cases. The red line represents the ideal output distribution, where all 64 states are well-separated and evenly distributed between 0 and 1. In contrast, the blue bars represent the experimentally obtained outputs for a non-optimized LHS pulse condition, which exhibits significant overlapping between states and non-monotonic current values, leading to poor encoding performance. This comparison highlights the critical role of optimizing pulse parameters in achieving clear separability between final states for different sequences, which is essential for reliable encoding and downstream neuromorphic computation [23].

Together, these results demonstrate how the TFT responds to different pulse sequences and why careful optimization of pulse parameters is essential for high-resolution multi-bit state encoding. The nonlinear current evolution and history-dependent accumulation effects are key physical mechanisms enabling this functionality. Moreover, the interplay among the five input parameters creates a high-dimensional, coupled space that is difficult to optimize by intuition or trial-and-error experimentation, making systematic optimization indispensable.

**Optimization of the Output State Separation via Bayesian Optimization**

To enhance the distinguishability of the 64 output states for 6-bit binary pulse sequences, we implement the previously described BO framework to efficiently explore the five-dimensional parameter space governing pulse conditions -- pulse period, base gate voltage, gate pulse amplitude, drain voltage, and duty cycle. Figure 4(a) shows the learning curve of the optimization process, where the x-axis represents the index of each tested condition and the y-axis shows the experimentally measured log(nDoS). The first 20 data points are generated using LHS to broadly cover the design space, while the remaining 25 are selected by the Bayesian optimizer over 5 rounds of 5 conditions each. For every condition, the average log(nDoS) over three measurements is plotted, along with error bars representing the standard deviation. The predictions from the GPR model (empty squares in Figure 4(a)) are also included to show the difference between model and experimental results.

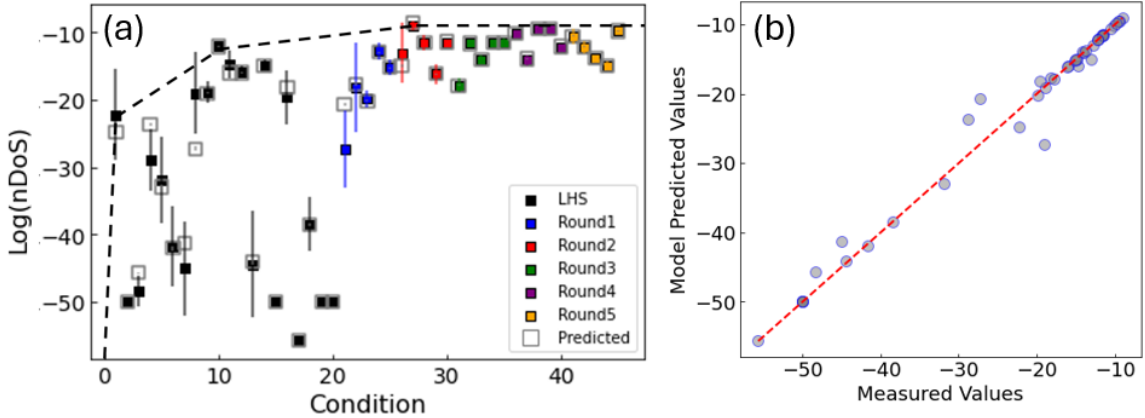

*Fig. 4 (a) Evolution of the log(nDoS) as a function of the evaluated experiment condition. Solid squares represent the experiments; each round is color coded. Open squares are the model-predicted values. The black dashed line is the highest log(nDoS) up to that point of optimization. (b) The predicted vs. experimentally measured log(nDoS) values for all evaluated conditions, demonstrating the high accuracy of the GP model.*

The initial LHS conditions produce widely ranging log(nDoS) values. The worst is approximately – 55, indicating inconsistent and poor separation among the 64 output states. Also note that the experimental and model-predicted values can differ by several orders of magnitude in the LHS conditions. In contrast, the BO-guided rounds rapidly identified high-performing regions of the parameter space. The agreement between model and experimental results also improves as the learning progresses, with the last 3 rounds of active learning showing excellent agreement between experiments and model. Furthermore, the standard deviations of the measured results are reduced in the later rounds. The best-performing condition was found in round 2 (condition #27), with a log(nDoS) of – 8.97, representing a substantial improvement over the LHS baseline. This optimal configuration corresponds to a pulse period of 66 ms, base gate voltage of 1.1 V, gate pulse amplitude = 1.9 V, $V_{DS}$ = 1 V, and duty cycle = 72%. Figure S3 shows the transient current responses as a function of time for all 64 input combinations of the 6-bit encoding, measured using this pulse condition. These time-resolved measurements demonstrate the distinct dynamic behavior of each binary sequence, confirming the effective separation of states achieved with the optimized pulse parameters. The complete list of all 45 pulse conditions, including their parameter values and nDoS results, is provided in the Supplementary Information (Table S1).

The accuracy of the model guiding the search is visualized in Figure 4(b), which compares predicted versus measured log(nDoS) for all conditions. The discrepancy between experimental and model-predicted values is smaller for the data points at the upper right, with log(nDoS) values between -10 and -20. This is because these data points are acquired in the later BO rounds. The excellent agreement confirms the effectiveness of the model in driving the optimization with minimal experimental overhead greatly reducing the number of required experiments, while ensuring good distinguishability among output states. To further illustrate the exploration behavior of the optimization process, 2D contour plots of the predicted log(nDoS) landscape for each parameter pair, along with all the measured points color-coded for each round, are included in the Supplementary Information (Figure S4). These plots reveal how the acquisition function ($q$UCB) strategically selects new conditions based on both expected performance and uncertainty. These contour plots highlight the regions in the parameter space where the predicted nDoS is maximized, indicating the optimal combinations of pulse parameters that the model identified as most promising during the search.

**Correlation between 4-bits and 6-bits models**

Since evaluating 6-bit encoding is time-consuming, we test whether a simpler 4-bit encoding model can inform or guide the optimization of the more complex 6-bit case. We conduct a parallel BO process using identical experimental conditions and pulse parameter ranges. Both models were trained on the same 45 experimental data points: 20 initial conditions generated via LHS and 25 additional points obtained through five rounds of BO active learning. For each condition, the log(nDoS) was computed independently for the two encoding models. The contour plots for the 4-bit model are shown in Figure S5. Figure S6 presents (a) the learning curve and (b) the parity plot comparing predicted versus experimental log(nDoS) values for the 4-bits model.

Despite the significant difference in encoding complexity—16 possible binary states in the 4-bit case vs. 64 in the 6-bit case—a strong correlation was observed between the predicted log(nDoS) values of the two models. Specifically, log(nDoS) values are much lower in the 6-bit case (ranging from -9 to -55) because the higher number of output states requires finer separation, even minor deviation in the final states can lead to significant penalties in log(nDoS), resulting in more negative values compared to the 4-bit model (from 0 to –8). This highlights the increased sensitivity and stricter requirements as the number of bits increases. To systematically explore the parameter space, we construct a uniform grid by discretizing each input parameter into 10 equally spaced values within their respective experimental ranges, resulting in 100,000 unique parameter combinations. As shown in Figure 5, a scatter plot comparing predicted log(nDoS) for 5,000 randomly selected points from this grid reveals a great similarity between the two models, with a Pearson correlation coefficient of 0.77. This result indicates that conditions leading to good output separation in the 4-bit model should generally also yield clear separation in the 6-bit scenario.

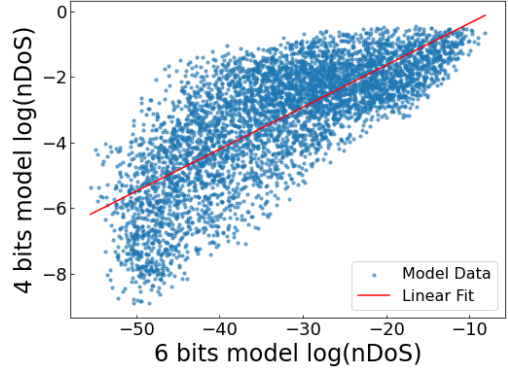

*Fig. 5 Scatter plot of log (nDoS) values predicted by the 4-bit model versus 6-bit model for 5,000 randomly sampled points from the full 100,000 points by discretizing each input variable in 10 equally spaced intervals. The solid red line represents the linear fit to the data, highlighting the positive correlation between the two models.*

Indeed, the best 4-bit condition achieved a log(nDoS) of –0.59 (nDoS = 0.25), whereas the optimal 6-bit condition required a much lower value of – 8.97 (nDoS = 1.1x10$^{-9}$) due to the greater number of required state distinctions and tighter spacing between output levels. This underscores the fact that as the number of bits increases, the output will be more susceptible to overlapping due to noise and therefore requires more precise control over pulse parameters. For the 6-bits optimum, the pulse period is 66 ms, base gate voltage 1.1 V, gate pulse amplitude 1.9 V, drain voltage 1 V and duty cycle of 72%. In contrast, the best 4-bits condition has a slightly shorter pulse period of 51 ms, a higher base gate voltage of 1.3 V, the same gate pulse amplitude and drain voltage, and higher duty cycle of 90%. Although the parameters are broadly similar, these differences highlight the finer tuning needed for 6-bit encoding due to its increased complexity. Remarkably, our 4-bit nDoS value of 0.25 is higher than those reported in previous works, including 0.10 for an IGZO TFT[4] and 3 x 10$^{-3}$ for a SnO TFT[25], further demonstrating the effectiveness of using BO strategy in enhancing the output state separation.

**Demonstration in Encoding Motion Images**

To validate the practical utility of the optimized multi-bit pulse encoding in a temporal data processing task, we implemented a simple spatiotemporal processing demonstration. The goal was to encode the movement of an object, in this case, a car moving across six sequential image frames, using the 6-bit pulse output of the TFT.

For six frames that capture a car moving from the lower left to the upper right of the image (Fig. 6(a)), we extract the binary values of the pixels from each frame, generating a 6-bit binary sequence that represents its temporal evolution. This sequence is then applied to the gate terminal under all the input conditions, and the final output current after each full sequence is used as a compressed representation of the motion of each pixel. We show the results from three pixels in Fig. 6(a). If the output is ideal, i.e., evenly spaced current states, the composite motion image is shown in Fig. 6(b), representing the best possible encoding performance. To evaluate the impact of pulse condition quality on encoding fidelity, we show three examples: i) a poor-performing condition randomly selected in the LHS batch (condition 2), ii) the optimal condition obtained via 6-bit BO model (condition 27), iii) the best-performing 4-bit condition applied directly to the 6-bits task (Condition 39). The normalized current vs # of pulses with the nDoS value and the reconstructed motion images for each of these pulse conditions are shown in Fig. 6(c). Visually, the poor-performing condition (left column in Fig. 6(c)) captures the general direction of the motion car; however, the reconstructed image appears saturated due to insufficient state separation, and the time sequence is lost as consecutive frames appear nearly identical. In particular, even low-value sequences such as 000000 yield relatively high normalized current levels, resulting in clustering of multiple states near the upper end of the range and a reduced dynamic contrast. On the other hand, the optimized 6-bit condition generate an image similar to using the ideal case, due to the well-separated current states and the highest nDoS (middle column in Fig. 6(c)). Remarkably, the optimized 4-bit condition performs the 6-bit task with almost indistinguishable results compared to the 6-bit optimum (right column in Fig. 6(c) right).

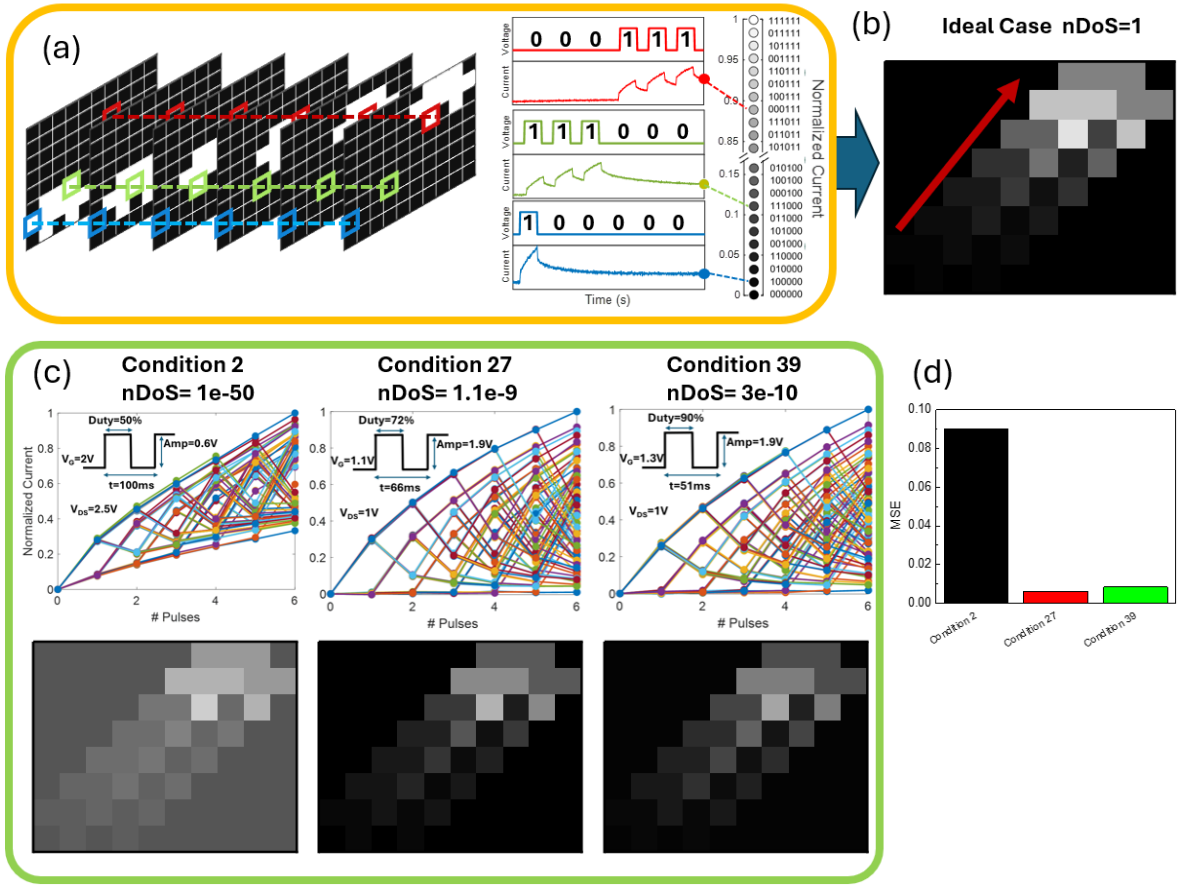

*Fig. 6 Demonstration of temporal encoding and decoding of a binary moving object using 6-bit pulse sequences. (a) Six sequential image frames showing the car movement. Three selected pixels are exemplified with their corresponding 6-bit pulse sequences and resulting normalized current values. (b) The composite motion image encoded using the ideal output result where all states are evenly spaced. (c) nDoS extracted from normalized current versus number of pulses for three pulse conditions: a poor-performing condition, the optimized 6-bit condition, and the optimized 4-bit condition applied to 6-bit encoding with their visualization of the reconstructed car movement for each condition. (e) Mean squared error (MSE) quantifying the reconstruction accuracy for the three conditions, demonstrating superior fidelity of the optimized conditions.*

To quantify the performance on motion image capture using different pulse conditions, we show the mean squared errors (MSE) between the reconstructed motion images (bottom row in Figure 6(c)) and the one made using the ideal output (Fig. 6(b)) in Figure 6(d). The MSE for poor condition yields a high value of 0.09, indicating the two images are substantially different. In contrast, the MSE for optimized 4-bit condition (0.0084) was comparable to that of the optimized 6-bit condition (0.0063), demonstrating that optimization based on 4-bit models can successfully predict high-performing pulse conditions for 6-bit encoding tasks. These findings underscore the possibility of using simplified models to accelerate hardware optimization for neuromorphic computing, as the strong agreement between the 4-bit-derived and 6-bit-optimal conditions suggests that the underlying device physics and nonlinear response are already well captured by the lower-complexity model.

**SHAP Analysis of Feature Importance**

To elucidate the factors driving the predictive performance of our BO model, we performed an interpretability analysis based on SHAP values[29,37]. SHAP quantifies the contribution of each input feature to the model's predicted outcome by measuring the change in prediction when varying that input values[38]. The magnitude of the SHAP value reflects the importance of each input variable, while the sign indicates whether the variable positively or negatively influences the predicted nDoS. A positive SHAP value means that higher values of the corresponding input parameter tend to increase the predicted nDoS, whereas a negative value suggests a decrease. Values close to zero imply a negligible effect on the model output.

Figure 7 summarizes the SHAP analysis for the 6-bit model, showing the mean absolute SHAP value of each parameter, which reflects its average impact on the model's predictions. Among the five parameters studied, the pulse gate amplitude emerged as the most influential factor, followed by the drain voltage $V_{DS}$, the duty cycle, the pulse period, and finally the base gate voltage. This ranking suggests that optimizing the gate pulse amplitude and the drain voltage has the greatest effect on achieving high output separation, while the gate voltage plays a secondary role.

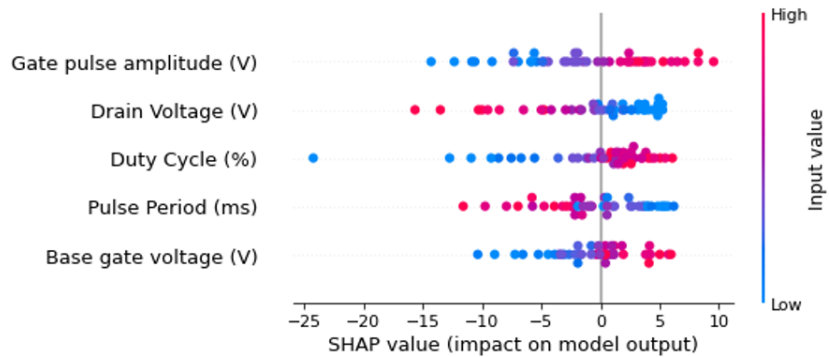

*Fig. 7 SHAP summary plot showing the relative impact of the five pulse parameters on the degree of separation in the Bayesian Optimization model*

The detailed SHAP dependence plots for each pulse parameter are provided in the Supplementary Information (Fig. S7). These plots reveal that higher gate pulse amplitude, base gate voltage, and duty cycle tend to increase the state separation, as evidenced by their SHAP values increasing with the parameter value. Conversely, increasing the drain voltage and pulse period will decrease the predicted nDoS value, suggesting that lower values of these parameters favor better state distinguishability.

Interestingly, the best-performing condition (#27) shows excellent agreement with the SHAP trends for the most influential parameters namely, high gate pulse amplitude and low drain voltage while exhibiting suboptimal or intermediate values for the remaining parameters. This observation reinforces the conclusion that a small subset of parameters dominates the device response, while others play a more secondary role.

These insights not only validate the relative importance of each parameter in the device's physical response but also provide valuable guidance for prioritizing variables in future optimization campaigns. The ability to identify key parameters that most strongly affect the neuromorphic performance of the TFT highlights the usefulness of combining machine learning optimization with interpretable AI tools such as SHAP.

**CONCLUSIONS**

In this work, we demonstrate a systematic and data-driven approach to optimize multi-bit pulse encoding in sol–gel processed $In_2O_3/Al_2O_3$ thin-film transistors for temporal data processing applications. Leveraging the inherent hysteresis and short-term memory effects of the TFTs, we implement a Bayesian optimization (BO) framework to identify optimal pulse parameters that maximize output separation, represented by log(nDoS) value, among 64 distinct states in a 6-bit encoding task. The BO process efficiently explores a five-dimensional input space -- pulse period, base gate voltage, gate pulse amplitude, drain voltage, and duty cycle -- achieving an optimal 6-bit configuration with clear state distinction and strong agreement with the ideal encoding behavior.

Additionally, we show that optimizing a simpler 4-bit model can serve as a practical and computationally efficient surrogate for guiding 6-bit optimization. The strong correlation observed between the log(nDoS) landscapes of the 4-bit and 6-bit models, and the comparable performance of the best 4-bit-derived condition when applied to the 6-bit task, validates this strategy. This approach reduces the experimental and computational burden associated with direct high-bit optimization, offering a scalable pathway for the design of complex neuromorphic systems without compromising performance.

Finally, we demonstrate the functional impact of the optimized conditions through a spatiotemporal image-encoding task, in which the optimized 6-bit and 4-bit-derived conditions produce accurate and high-fidelity reconstruction of object motion sequences. These results show that solution-processed oxide TFTs are potentially a flexible and reconfigurable platform for implementing advanced spatiotemporal computing tasks. An interpretable analysis using SHAP further highlights the relative importance of pulse parameters, providing valuable insight into the physical and operational factors driving device performance.

This work establishes that BO provides a powerful tool to identify optimal operating condition of a physical reservoir, especially for complex multi-bit tasks with high-dimensional inputs. The approach demonstrated here is not limiting to TFTs and can be applied to optimizing other physical reservoir implementations across different material platforms.

## Methods

### Device Fabrication

The preparation of the sol-gel precursor solutions has been reported previously[36]. Briefly, the $Al_2O_3$ solution was prepared by dissolving 0.4 M of aluminum nitrate nonahydrate (Al $(NO_3)_3 \cdot 9H_2O$) in 2-methoxyethanol (2-MOE). The solution was stirred overnight at 80°C prior to deposition. The 0.2 M $In_2O_3$ precursor solution was obtained by dissolving indium nitrate hydrate ($In(NO_3)_3 \cdot 3H_2O$) in 2-MOE for 24 hrs with stirring. After adding 0.2 M acetylacetone and 0.2 M ammonium hydroxide, the final mixture was stirred for an additional 24 hrs at room temperature[39].

For the fabrication of MIM capacitors and TFT devices, the $Al_2O_3$ solution was filtered through 0.2 µm polytetrafluoroethylene filter and deposited onto precleaned ITO-coated (100 nm) glass substrates by spin-coating at 3000 rpm for 30 seconds. The $Al_2O_3$ dielectric films were dried at 85°C for three minutes and annealed on a hot plate at 250 °C for 1 hour, yielding a dielectric thickness of 80 nm. The $In_2O_3$ semiconductor layer, with a final thickness of 30 nm, was deposited onto the $Al_2O_3$ dielectric layer using the same spin-coating, drying and annealing conditions[36].

To define the $In_2O_3$ semiconductor patterns, an invar shadow mask was placed directly onto the spin-coated $In_2O_3$ precursor layer. The sample was then exposed to UV–ozone treatment for 10 minutes to initiate conversion by breaking down nitrate in the exposed precursor film to form an indium hydroxide network resistant to the developing solution (methanol: deionized water: acetic acid in a volume ratio of 15:5:1)[40]. This process selectively removed the unexposed regions of the film, yielding a well-defined semiconductor pattern. MIM capacitors were fabricated in the areas where the $In_2O_3$ was removed. To complete the device fabrication, 100 nm of aluminum contacts were deposited through a shadow mask using thermal evaporation. The channel dimensions of the resulting TFTs were 100 µm in length and 500 µm in width.

### Electrical Characterization

The electrical properties of the $Al_2O_3$ films were evaluated using MIM capacitor structures. Leakage current measurements were performed with a Keithley 4200A-SCS parameter analyzer connected to a probe station. *C–f* characteristics were also measured on the MIM devices using an Agilent 4284A precision LCR meter over a frequency range of 100 Hz to 1 MHz. The electrical characterization of the TFTs was carried out by measuring the transfer curves at room temperature and under dark conditions using a Keithley 4200A-SCS semiconductor parameter analyzer.

Pulsed voltage measurements were performed using a custom setup designed to evaluate the transient response of the TFT devices under dynamic electrical stimulation (Fig. S8). An Agilent 81110A pulse generator was used to apply gate voltage pulses and drain voltage. The source current is measured using a Stanford Research Systems SR570 low-noise current preamplifier. A Tektronix DPO 5104 digital oscilloscope is used for acquisition and analysis of time-dependent current response with a data acquisition rate of 2 ns. To ensure consistent initial current before each pulse sequence, the device was allowed to relax at the base gate voltage for 300 ms before starting the pulse sequence, followed by the application of a brief (3 s) negative bias stress ($V_G$ = -2 V and $V_{DS}$ = -3 V) at the end of each pulse sequence to remove residual charge and restore the baseline state.

**Bayesian Optimization Procedure**

To identify the optimal electrical pulse conditions that maximize the distinguishability of output current states in a neuromorphic TFT, we implemented a BO framework using GPR. The optimization aimed to maximize the log(nDoS) across 64 possible 6-bit combinations.

The five inputs, their explored ranges and increments are summarized in the following table. The total number of distinct input conditions, i.e., with at least one input with a different value, is 1.28 x 10$^7$, which is impossible to examine by traditional methods.

| Input | Lower limit | Upper limit | increment | # of values |
|---|---|---|---|---|
| Pulse period (ms) | 30 | 100 | 1 | 71 |
| Base gate voltage (V) | 1 | 2 | 0.1 | 11 |
| Gate pulse amplitude(V) | 1 | 2 | 0.1 | 11 |
| Drain voltage (V) | 1 | 3 | 0.1 | 21 |
| Duty Cycle (%) | 20 | 90 | 1 | 71 |

An initial set of 20 input conditions was generated using LHS to ensure uniform coverage of the parameter space. Each condition was applied to the TFT device using a 6-bit binary pulse train, and the resulting transient source current was recorded for each of the 64 binary combinations. For each sequence, the current values after each pulse (extracted at specific time points) were used to construct a final current level, normalized to the final current obtained after the sixth pulse of the 111111 sequence, in order to obtain current values from 0 to 1. The separation between output states was quantified using the DoS, defined as:

$$DoS = \prod_{i=1}^{n} \frac{c_i}{I_{DS_{max}} - I_{DS_{min}}} \qquad (1)$$

where *n* represents the number of distinguishable states, and $c_i$ represents the current difference between adjacent states, normalized to the dynamic range of the device defined by $I_{DS_{max}} = 1$ and $I_{DS_{min}} = 0$ [4,25]. To enable a meaningful comparison with an ideal, perfectly uniform output distribution, the DoS value for a given pulse condition was normalized to the DoS of the ideal case, resulting in the nDoS:

$$nDoS = \frac{measured\ DoS}{ideal\ DoS} \qquad (2)$$

For reference, the value of the ideal DoS is 2.28x10$^{-18}$ for 4-bit encoding and 4.38x10$^{-114}$ for 6-bits encoding, reflecting the increasingly strict separation requirements as the number of output states increases.

Due to the wide range of nDoS values, the logarithm of nDoS, denoted as log(nDoS), was used as the objective function during optimization.

The GPR model was implemented using the BoTorch library in python, employing a Matern 5/2 kernel with Automatic Relevance Determination to allow for separate length scales across the five input dimensions. A FixedNoiseGP model was used to incorporate measurement uncertainty and the batch upper confidence bound acquisition function guided the optimization. This function is defined as:

$$qUCB = \mu(x) + \beta * \sigma(x) \qquad (3)$$

Where *μ(x)* and *σ(x)* represent the mean prediction and the predictive uncertainty scaled by the hyperparameter β, which is set to 1.0 to ensure a balance between exploration and exploitation during the optimization[41].

At each iteration, the model proposed a batch of five new conditions (*q* = 5), which were experimentally evaluated to obtain additional log(nDoS) values. Combining the new results with existing results were used to retrain the model. The optimization loop (sampling → evaluation → training → model update) was repeated until the log(nDoS) converged and no longer showed significant improvement.

The final output of the process was the pulse condition yielding the highest log(nDoS), selected as the optimal configuration for high-fidelity 6-bit encoding.

## Funding declaration

This work was supported by the Air Force Office of Scientific Research under award number FA2386-24-1-4041, the National Science Foundation Award CMMI-2135203, and the Taiwan National Science and Technology Council Award 113-2124-M-006-008-MY3. J. M. gratefully acknowledges funding from the Fulbright García Robles fellowship for postdoctoral research. J.W.P.H. acknowledges the support of the Texas Instruments Distinguished Chair in Nanoelectronics.

## Acknowledgements

The authors thank W. Vandenberghe for helpful suggestions.

## Author contributions

J.H. and J.-S.C. conceived the idea and contributed to the conceptualization of the project. J.M. wrote the manuscript, set up the system for pulsed measurements, and, together with B.D., designed and carried out the experiments, developed the methodology, and fabricated the devices. Y.-C.C. provided critical information on performing measurements correctly. J.M. and B.D. developed the Bayesian optimization code, and W.X. implemented the SHAP analysis code. All authors analyzed data and provided scientific input. J.H. supervised and directed the project and secured funding. All authors reviewed, edited, and approved the final version of the manuscript.

## Competing interest

The authors declare no competing interests.

# Supplementary Information

# Bayesian Optimization of Multi-Bit Pulse Encoding in $In_2O_3/Al_2O_3$ Thin-film Transistors for Temporal Data Processing

Javier Meza-Arroyo[a], Benius J. Dunn[a], Weijie Xu[a], Yu-Chieh Chen[b], Jen-Sue Chen[b], Julia W.P. Hsu[a]

[a] Department of Materials Science and Engineering, The University of Texas at Dallas, Richardson, Texas 75080, USA

[b] Department of Materials Science and Engineering, National Cheng Kung University, Tainan 70101, Taiwan.

| Condition | Pulse period (ms) | Base gate voltage (V) | Gate pulse amplitude (V) | Drain Voltage (V) | Duty Cycle (%) | Log (nDoS) 4 bits | nDos 4 bits | Log (nDoS) 6 bits | nDos 6 bits |
|---|---|---|---|---|---|---|---|---|---|
| 1 | 40 | 1 | 1.6 | 2.4 | 80 | -2.28 | 5.25E-03 | -22.21 | 6.17E-23 |
| 2 | 100 | 2 | 0.6 | 2.5 | 50 | -4.51 | 3.09E-05 | -50.00 | 1.00E-50 |
| 3 | 80 | 1.2 | 1 | 1.4 | 70 | -2.25 | 5.62E-03 | -48.33 | 4.68E-49 |
| 4 | 50 | 1.6 | 0.6 | 1.8 | 70 | -2.94 | 1.15E-03 | -28.87 | 1.35E-29 |
| 5 | 70 | 1.8 | 0.8 | 2 | 40 | -4.13 | 7.41E-05 | -31.84 | 1.45E-32 |
| 6 | 80 | 1.8 | 0.4 | 1.6 | 30 | -7.80 | 1.58E-08 | -41.70 | 2.00E-42 |
| 7 | 60 | 1.4 | 1 | 1.6 | 30 | -3.12 | 7.59E-04 | -45.04 | 9.12E-46 |
| 8 | 90 | 1.4 | 1.2 | 1.4 | 50 | -1.81 | 1.55E-02 | -19.02 | 9.55E-20 |
| 9 | 30 | 2 | 1 | 2.6 | 50 | -2.28 | 5.25E-03 | -18.86 | 1.38E-19 |
| 10 | 70 | 1.2 | 1.6 | 1 | 90 | -0.73 | 1.86E-01 | -12.08 | 8.32E-13 |
| 11 | 50 | 1.4 | 0.6 | 1.2 | 80 | -2.36 | 4.37E-03 | -14.67 | 2.14E-15 |
| 12 | 50 | 1.8 | 1.2 | 2.2 | 60 | -1.99 | 1.02E-02 | -15.93 | 1.17E-16 |
| 13 | 90 | 1.2 | 1.6 | 2.4 | 20 | -4.11 | 7.76E-05 | -44.37 | 4.27E-45 |
| 14 | 40 | 1.6 | 0.8 | 1.8 | 70 | -1.88 | 1.32E-02 | -14.86 | 1.38E-15 |
| 15 | 60 | 1 | 1.4 | 3 | 40 | -8.00 | 1.00E-08 | -50.00 | 1.00E-50 |
| 16 | 40 | 1.8 | 1 | 2 | 90 | -1.36 | 4.37E-02 | -19.64 | 2.29E-20 |
| 17 | 30 | 1.6 | 0.6 | 1.2 | 20 | -4.82 | 1.51E-05 | -55.70 | 2.00E-56 |
| 18 | 70 | 1.4 | 1 | 2.2 | 70 | -2.31 | 4.90E-03 | -38.46 | 3.47E-39 |
| 19 | 80 | 1.6 | 0.4 | 2.6 | 60 | -8.00 | 1.00E-08 | -50.00 | 1.00E-50 |
| 20 | 60 | 1.2 | 1.6 | 2.8 | 40 | -3.15 | 7.08E-04 | -50.00 | 1.00E-50 |
| 21 | 30 | 2 | 1 | 1.5 | 58 | -2.83 | 1.48E-03 | -27.33 | 4.68E-28 |
| 22 | 97 | 1.4 | 1.6 | 1 | 44 | -1.23 | 5.89E-02 | -18.16 | 6.92E-19 |
| 23 | 52 | 2 | 1 | 1.3 | 38 | -2.81 | 1.55E-03 | -19.88 | 1.32E-20 |
| 24 | 98 | 1.2 | 1.8 | 1 | 83 | -1.33 | 4.68E-02 | -12.78 | 1.66E-13 |
| 25 | 67 | 2 | 1 | 1.4 | 69 | -1.81 | 1.55E-02 | -15.10 | 7.94E-16 |
| 26 | 30 | 1.8 | 0.6 | 1.5 | 77 | -1.90 | 1.26E-02 | -13.03 | 9.33E-14 |
| 27 | 66 | 1.1 | 1.9 | 1 | 72 | -1.17 | 6.76E-02 | -8.97 | 1.1E-09 |
| 28 | 59 | 1.5 | 1.5 | 1 | 90 | -1.11 | 7.76E-02 | -11.55 | 2.82E-12 |
| 29 | 78 | 1.8 | 1.1 | 1 | 49 | -1.97 | 1.07E-02 | -16.18 | 6.61E-17 |
| 30 | 30 | 1.6 | 1 | 1.7 | 72 | -1.61 | 2.45E-02 | -11.44 | 3.63E-12 |
| 31 | 30 | 1.9 | 0.7 | 2 | 63 | -2.11 | 7.76E-03 | -17.84 | 1.45E-18 |
| 32 | 44 | 1.3 | 1.5 | 1.3 | 81 | -1.02 | 9.55E-02 | -11.63 | 2.34E-12 |
| 33 | 47 | 1.7 | 1 | 1.2 | 70 | -1.42 | 3.80E-02 | -14.05 | 8.91E-15 |
| 34 | 72 | 1.5 | 1.4 | 1 | 71 | -0.84 | 1.45E-01 | -11.57 | 2.69E-12 |
| 35 | 30 | 1.5 | 1 | 1.4 | 87 | -0.94 | 1.15E-01 | -11.53 | 2.95E-12 |
| 36 | 72 | 1.4 | 1.9 | 1 | 78 | -0.68 | 2.09E-01 | -10.14 | 7.24E-11 |
| 37 | 61 | 1.8 | 1.6 | 1 | 57 | -2.12 | 7.59E-03 | -14.11 | 7.76E-15 |
| 38 | 60 | 1.6 | 1.6 | 1 | 60 | -0.89 | 1.29E-01 | -9.39 | 4.07E-10 |
| 39 | 51 | 1.3 | 1.9 | 1 | 90 | -0.59 | 2.57E-01 | -9.53 | 3.00E-10 |
| 40 | 85 | 1.7 | 1.7 | 1 | 66 | -1.09 | 8.13E-02 | -12.13 | 7.41E-13 |
| 41 | 85 | 1.4 | 1.7 | 1 | 73 | -0.72 | 1.91E-01 | -10.54 | 2.88E-11 |
| 42 | 35 | 1.2 | 1.9 | 1 | 69 | -0.90 | 1.26E-01 | -12.15 | 7.08E-13 |
| 43 | 30 | 1.3 | 0.5 | 1 | 90 | -1.83 | 1.48E-02 | -13.75 | 1.78E-14 |
| 44 | 31 | 1.6 | 0.5 | 1 | 90 | -2.13 | 7.41E-03 | -15.00 | 1.00E-15 |
| 45 | 72 | 1.4 | 1.7 | 1 | 61 | -1.32 | 4.79E-02 | -9.81 | 1.55E-10 |

Table S1 Complete list of tested pulse parameter combinations with corresponding Degree of Separation results for 4-bits and 6-bits models

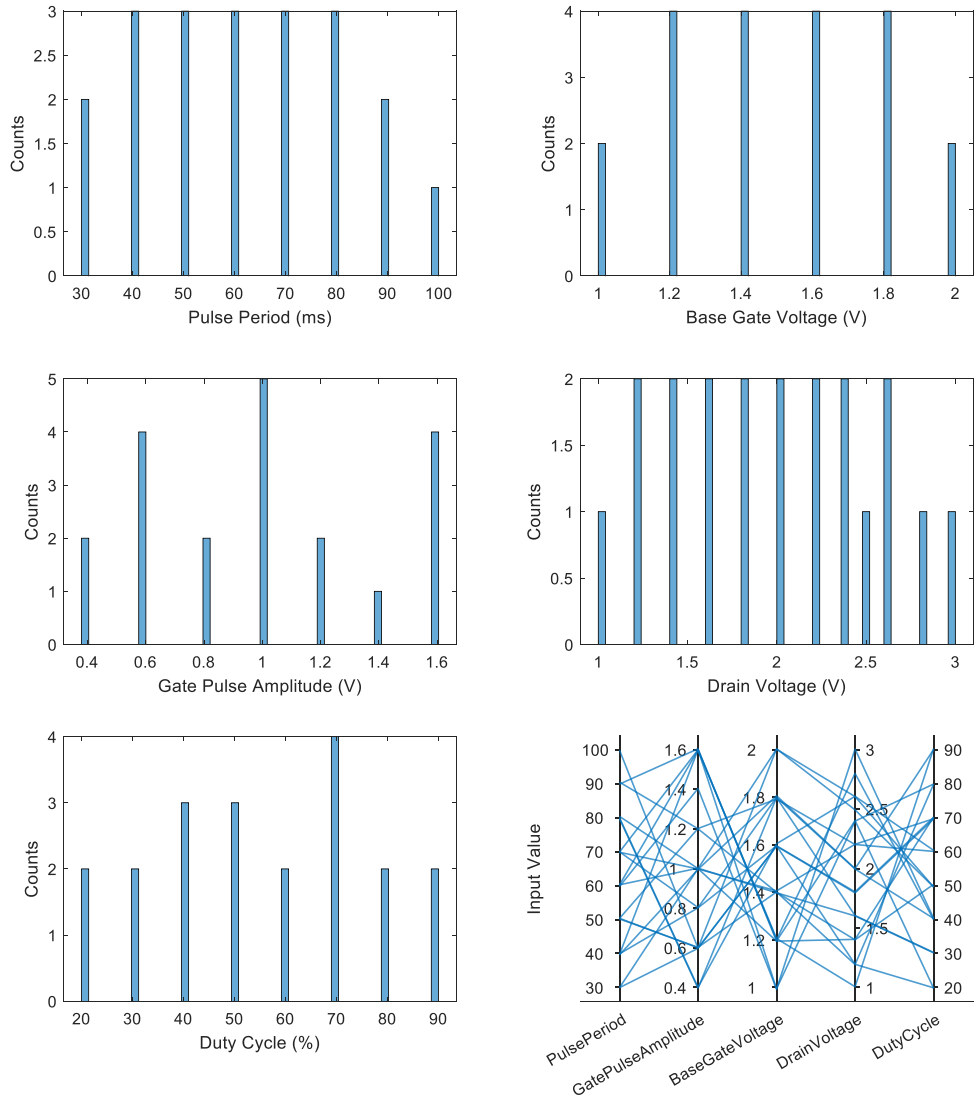

Figure S1 Distribution of the 20 initial input conditions generated by LHS across the five-dimensional pulse parameter space

To ensure reliable pulse-based encoding, the device must start with the same initial state for each pulse sequence. Figure S2(a), shows the transient drain current under multiple positive bias application at $V_G$ = 2 V and $V_{DS}$ = 3 V for ~4 seconds. The solid black curve corresponds to the first bias application, exhibiting a steady increase in current. After a 1-second zero-bias interval, the same positive biases were reapplied and the second bias response (red curve) was recorded, showing a higher current due to memory effect of the previous measurement. This process was repeated five times, with each subsequent bias yielding progressively higher current levels.

Following the fifth positive bias (yellow curve), a reset protocol was applied using negative biases ($V_G$ = –2 V, $V_{DS}$ = –3 V) for 3 seconds before another set of 5 positive bias measurements were repeated. Fig. S2(a) shows that the reset protocol returns the drain current to its initial value as shown by the black dashed curve matching the first bias of the previous set closely, indicating successful restoration of the device to its baseline state. This reset and measurement process was repeated three times in total, and in each case, the initial conditions were reproducible with negligible deviation.

To highlight the cumulative current increase during consecutive biases, Figure S2(b) plots the drain current at a fixed time point (indicated by the vertical red dashed line in Fig. S2(a)) vs the positive bias count within each set. The repeatability of current values across three sets of measurements confirms the reliability and effectiveness of the reset protocol. This procedure is critical to ensure consistent starting conditions for pulse-based encoding and repeatable device response across multiple trials.

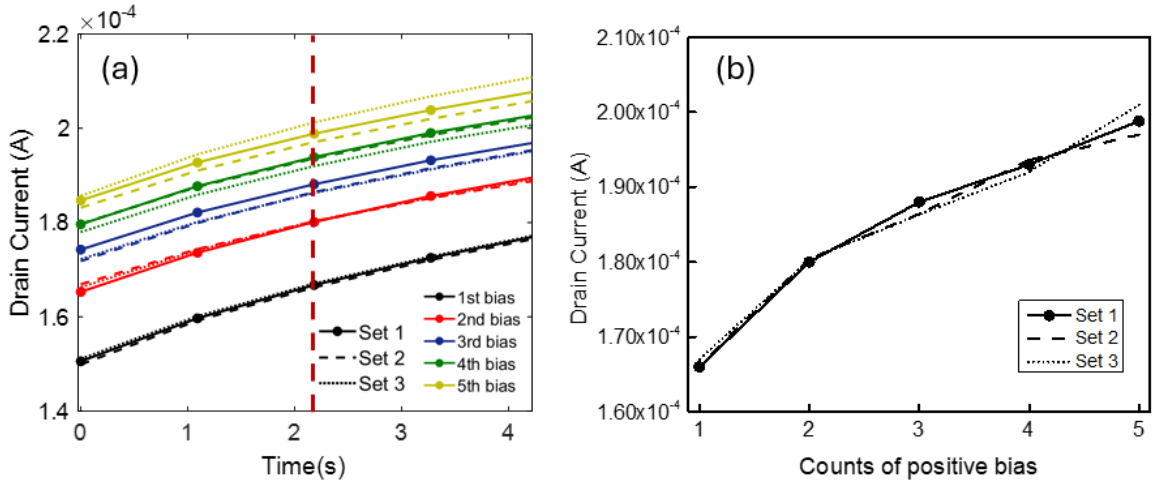

Figure S2 (a) Time dependence of current for three sets of positive bias measurements; in each set, the TFT is positively biased five times. In each set of measurements, the subsequent application of positive bias results in progressively higher current due to memory effects, illustrating the history-dependent behavior of the TFT. A reset protocol of negative bias is applied after each set, i.e., five positive bias applications, which restores the devices to its baseline states. (b) Drain current at a fixed time indicated by the red dashed line in (a) as a function of positive bias counts for the three sets of measurements.

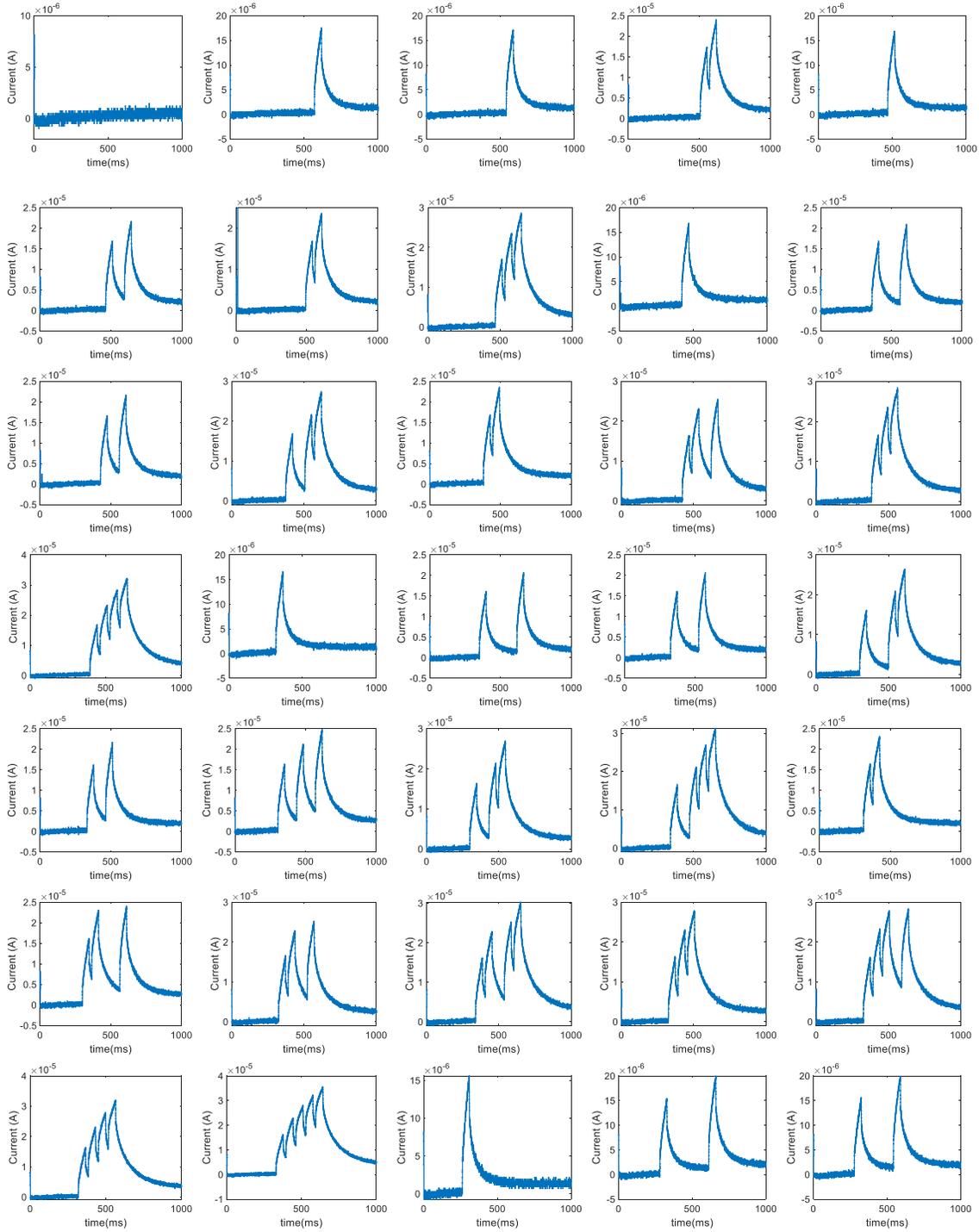

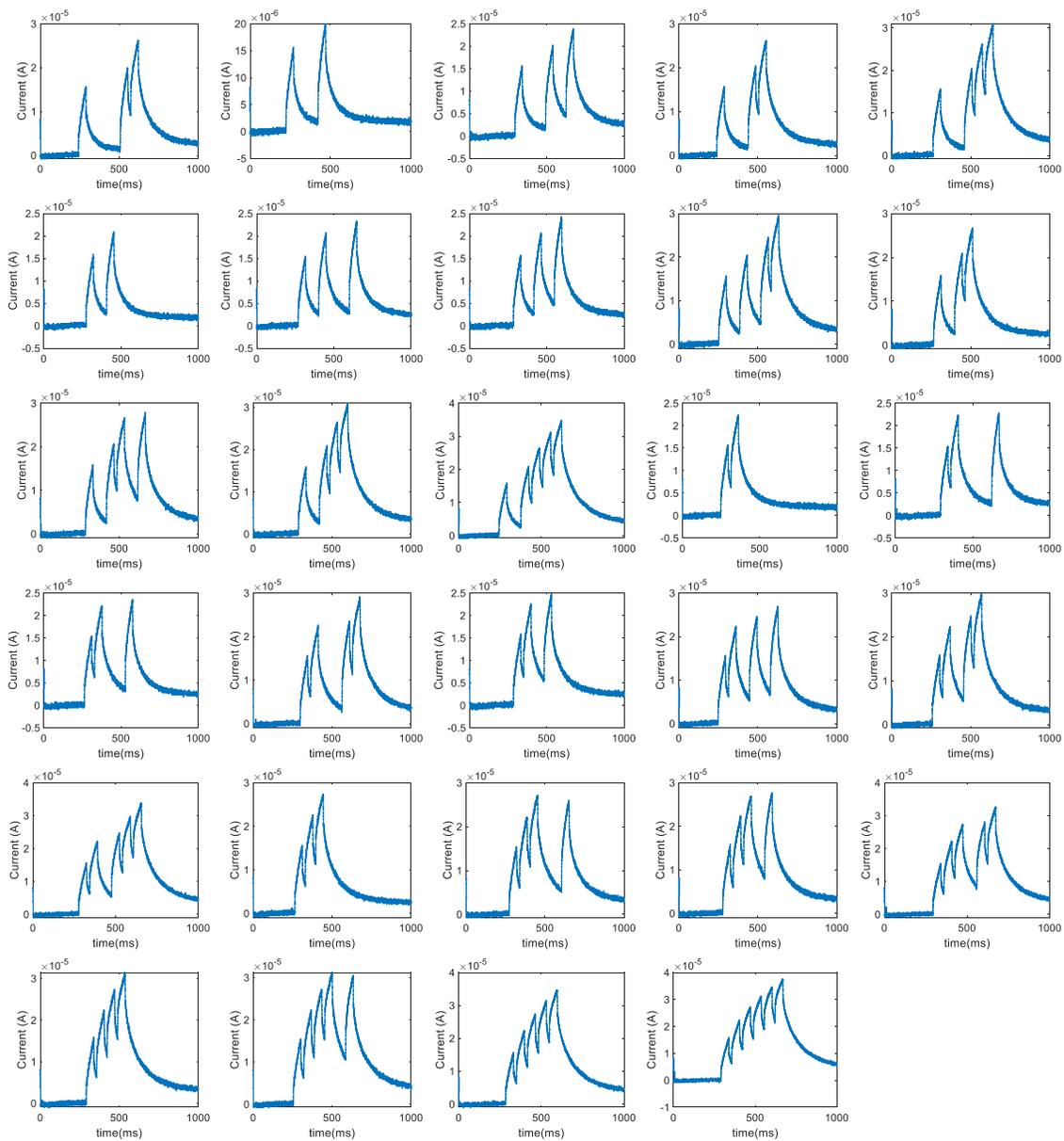

Figure S3 Transient drain current responses as a function of time for all 64 inputs combinations of the 6-bits encoding of condition 27

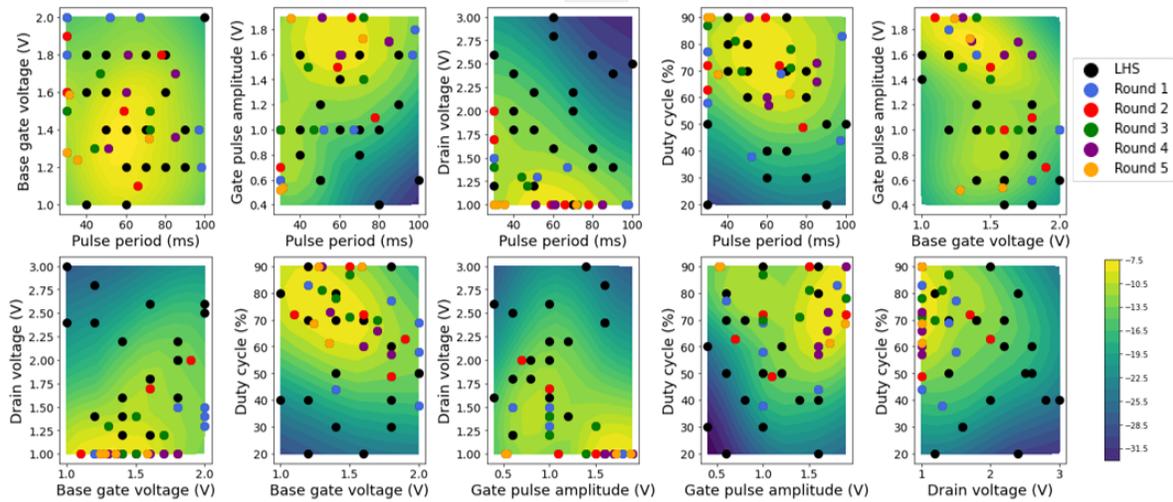

Figure S4 2D contour plots of the predicted log(nDoS) landscape for selected pairs of pulse parameters in the 6-bits model, showing LHS initial points and subsequent BO Rounds.

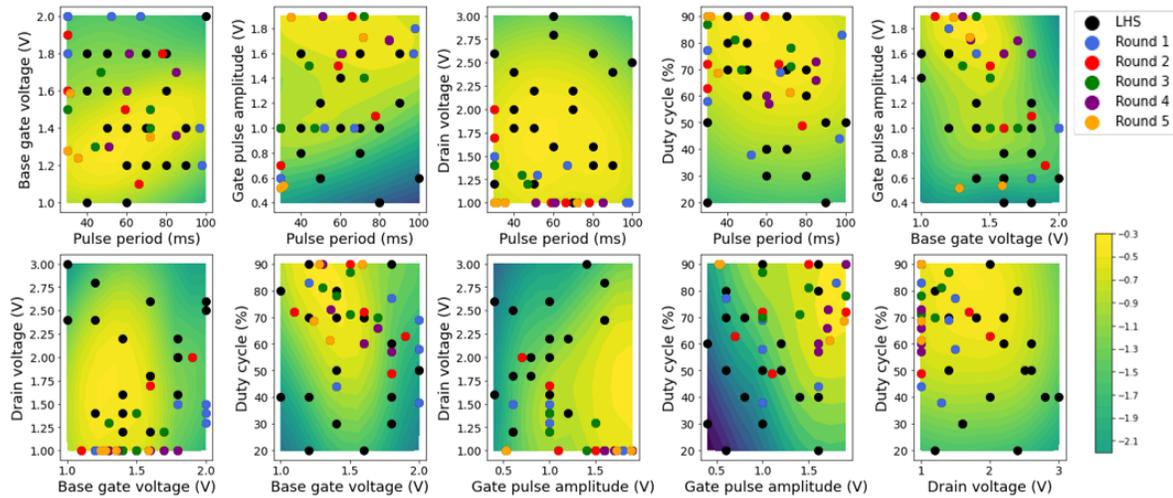

Figure S5 2D contour plots of the predicted log(nDoS) landscape for selected pairs of pulse parameters in the 4-bits model, showing LHS initial points and subsequent Bayesian Optimization Rounds

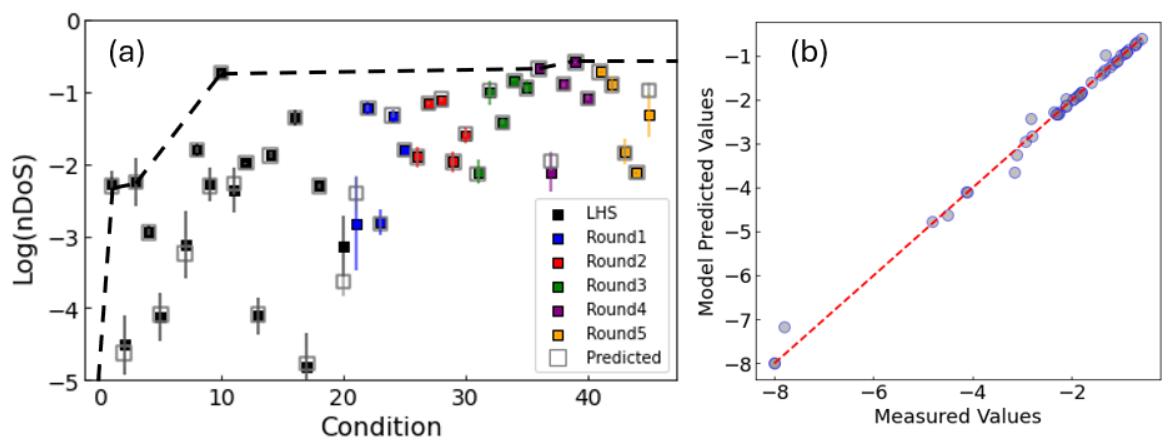

Figure S6 (a) Evolution of the log(nDoS) as a function of the evaluated experiment condition for 4-bit model. Parity plot comparing the predicted and experimentally measured log (nDoS) values for all evaluated condition, demonstrating the high accuracy of the 4 bits GP model.

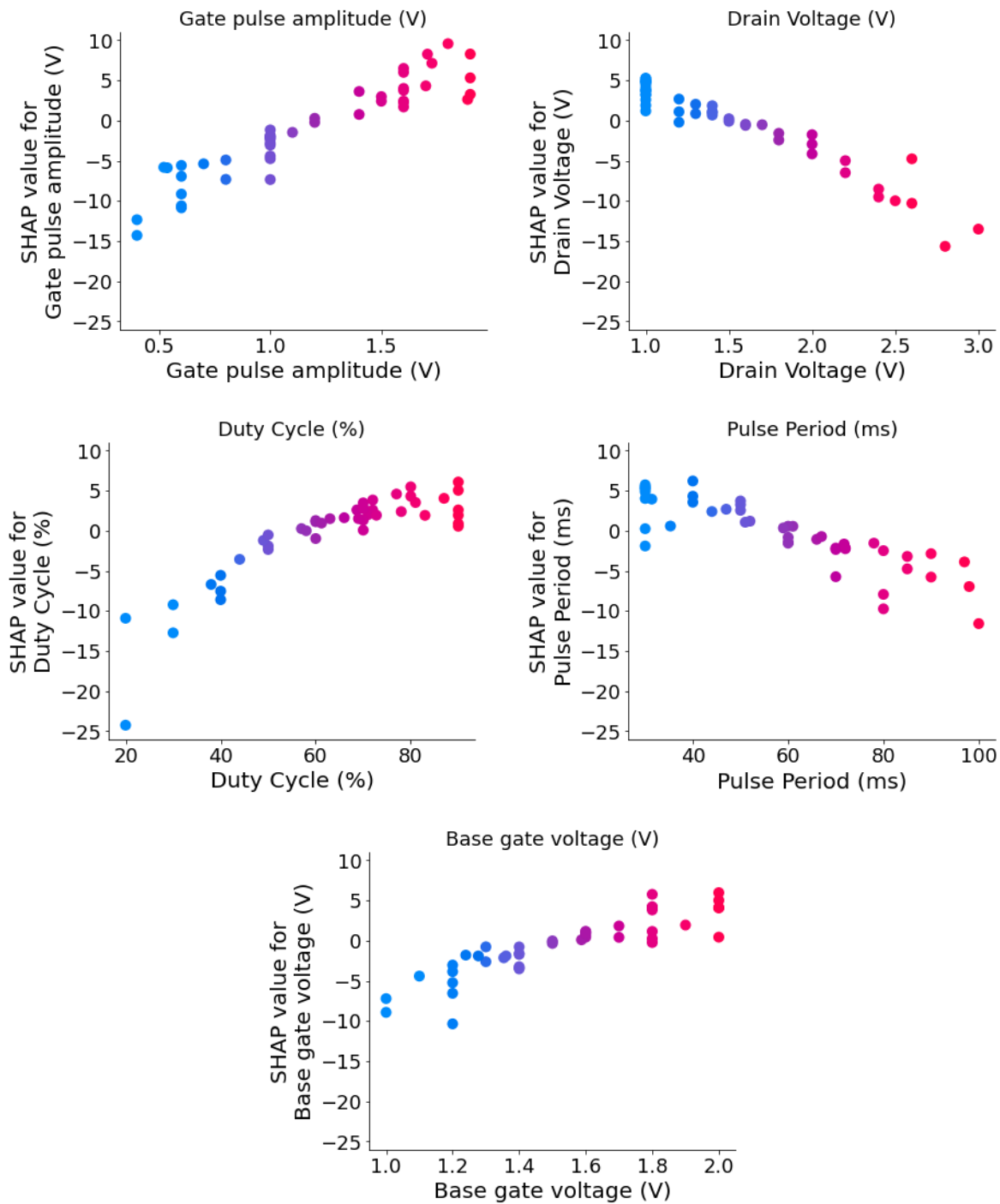

Figure S7 SHAP dependence plots for each of the five pulse parameters in the 6-bits model

Figure S8 shows the schematic of the experimental setup used to perform pulsed gate and drain measurements on the sol–gel processed TFTs. An Agilent 81110A pulse/pattern generator applied both gate ($V_G$) and drain ($V_{DS}$) voltage pulses. Channel 1 of the generator was connected to the gate terminal, while Channel 2 was connected to the drain and fixed to DC voltage. The source current was measured using a Stanford Research Systems SR570 low-noise current preamplifier, which converted the current into a voltage signal. This signal was then sent to a Tektronix DPO 5104 digital oscilloscope, which provided high temporal resolution (2 ns sampling rate).

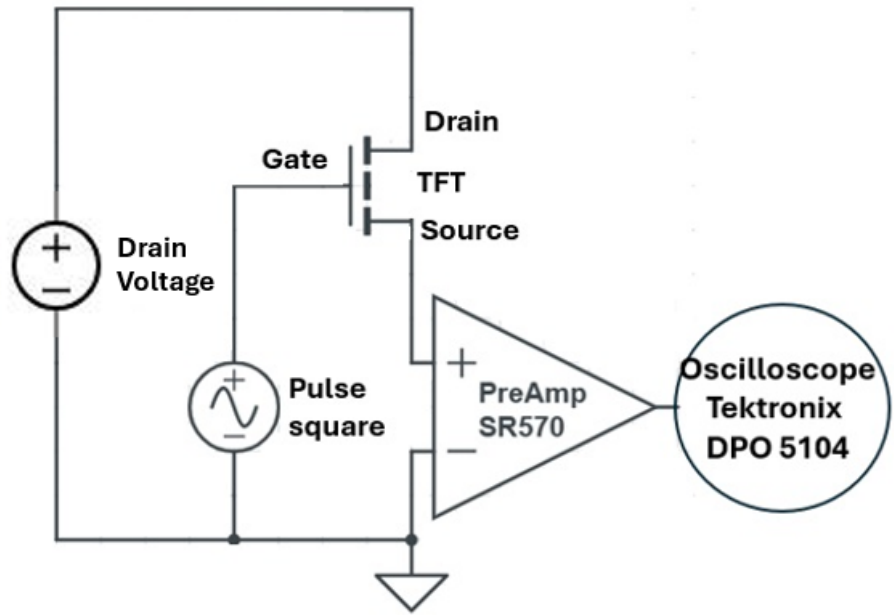

Figure S8 Circuit schematic of the measurement setup for pulse voltage experiments